\RequirePackage{fix-cm}
\documentclass[smallcondensed]{svjour3}     
\smartqed  
\usepackage{graphicx}
\usepackage{amssymb}
\usepackage{amsmath}
\usepackage{bm}
\usepackage{appendix}
\begin{document}

\title{A Numerical Method to Find the Optimal Thermodynamic Cycle in Microscopic Heat Engine
}

\titlerunning{Optimal Thermodynamic Cycle in Microscopic Heat Engine}        

\author{Rongxing Xu 
}

\authorrunning{Rongxing Xu} 

\institute{Rongxing Xu \at
               \email{xurongxing@keio.jp}\\
              Department of Physics, Keio University, 3-14-1 Hiyoshi, Yokohama 223-8522, Japan \\
              Mathematical Science Team, RIKEN Center for Advanced Intelligence Project (AIP), 1-4-1 Nihonbashi, Chuo-Ku, Tokyo 103-0027, Japan \\
}

\date{Received: date / Accepted: date}

\maketitle

\begin{abstract}
  Heat engines are fundamental physical objects to develop the nonequilibrium thermodynamics. The thermodynamic performance of the heat engine is determined by the choice of cycle and time-dependence of parameters. Here, we propose a systematic numerical method to find a heat engine cycle to optimize some target functions. We apply the method to heat engines with slowly varying parameters and show that the method works well. Our numerical method is based on the genetic algorithm which is widely applied to various optimization problems. 
\end{abstract}

\section{Introduction}\label{intro}
The last several decades have witnessed significant progress in nonequilibrium thermodynamics triggered by several important discoveries including the fluctuation theorem \cite{evans2002fluctuation} and nonequilibrium work theorem \cite{crooks1999entropy,jarzynski2004nonequilibrium}. Stochastic thermodynamics provides a robust framework in nonequilibrium thermodynamics focusing on the work, heat, and entropy production. Heat engines have been critical physical objects to develop nonequilibrium thermodynamics \cite{benenti2017fundamental}. Recently, there are many experimental works on heat engines. State-of-art experiments demonstrate microscopic scale heat engines using Brownian particles \cite{martinez2015adiabatic,martinez2016brownian}, single-qubit systems \cite{ronzani2018tunable}, and atomic-scale systems \cite{rossnagel2016single}. These are very crucial to understand the role of thermal and quantum fluctuations in small systems and the practical controllability of thermodynamic states. On the theoretical side, one of the important subjects is to figure out the robust relations between thermodynamic efficiency and output power. Universal aspects on the efficiency at the maximum power \cite{curzon1975efficiency,van2005thermodynamic,esposito2009universality} are important not only as of the fundamental physics but also practically for our daily life. Motivated by many realistic heat engines, the universal relations between efficiency and power have been discussed for classical systems \cite{brandner2015thermodynamics}, linear response regimes \cite{shiraishi2016universal}, and near quasi-static cycle \cite{brandner2020thermodynamic}.

To develop a further understanding of heat engines, another crucial problem is a methodology for controlling or optimization of thermodynamic states. Schmidle and Seifert have developed a method to find an optimization protocol to transform one particle distribution into another \cite{schmiedl2007optimal}. Later, the operating time required to achieve desired distribution has been intensively studied in terms of the speed limit \cite{shiraishi2018speed,funo2019speed}. However, we note that a systematic method to find an optimal cycle in a heat engine has not been seriously discussed. The problem here is whether there is a systematic approach to find a heat engine cycle to optimize some target function within a fixed available parameter region. Difficulties on this problem lie in the point that parameters in the heat engine cycles are continuous variables and hence the analytical approach is extremely difficult. In this paper, we present a systematic numerical approach to find the heat engine cycle. To discuss our method clearly, we focus on heat engines with slow parameter driving near quasi-static regime, where recently Brandner and Saito developed a theory on the relation between the efficiency and power \cite{brandner2020thermodynamic}. In their theory, they fix a cycle, and provide a geometrical framework to characterize the efficiency and power using the thermodynamic length ${\cal L}$ and quasi-static work ${\cal W}$. According to their theory, a large value of ${\cal W}/{\cal L}$ is beneficial for heat engine performance. In this paper, we present a systematic numerical method to find an optimal cycle to maximize the function ${\cal W}/{\cal L}$ for a given parameter space. 

A key idea in our work is to use the genetic algorithm that has been widely developed and applied to various optimization problems in computer science \cite{whitley1994genetic}. We show that this method works well in heat engine problems also. The convergence can be checked through a relevant error estimation. By developing the numerical method, we also obtain a physical insight for optimizing the cycles which explicitly shows that large power and high efficiency cannot be obtained simultaneously.

This paper is organized as follows. We first briefly introduce Brandner and Saito's work in section \ref{section: 2} and apply them on both microscopic heat engines as section \ref{section: 3}. We then illustrate the numerical method and its application to our problem in section \ref{section: 4}. The numerical results are discussed in section \ref{section: 5}.

\section{Heat engine with slow driving and main goal of this paper}\label{section: 2}

\subsection{Power-efficiency trade-off relation}
\label{sec2subsec1}

\begin{figure}[t]
\includegraphics[width=\textwidth]{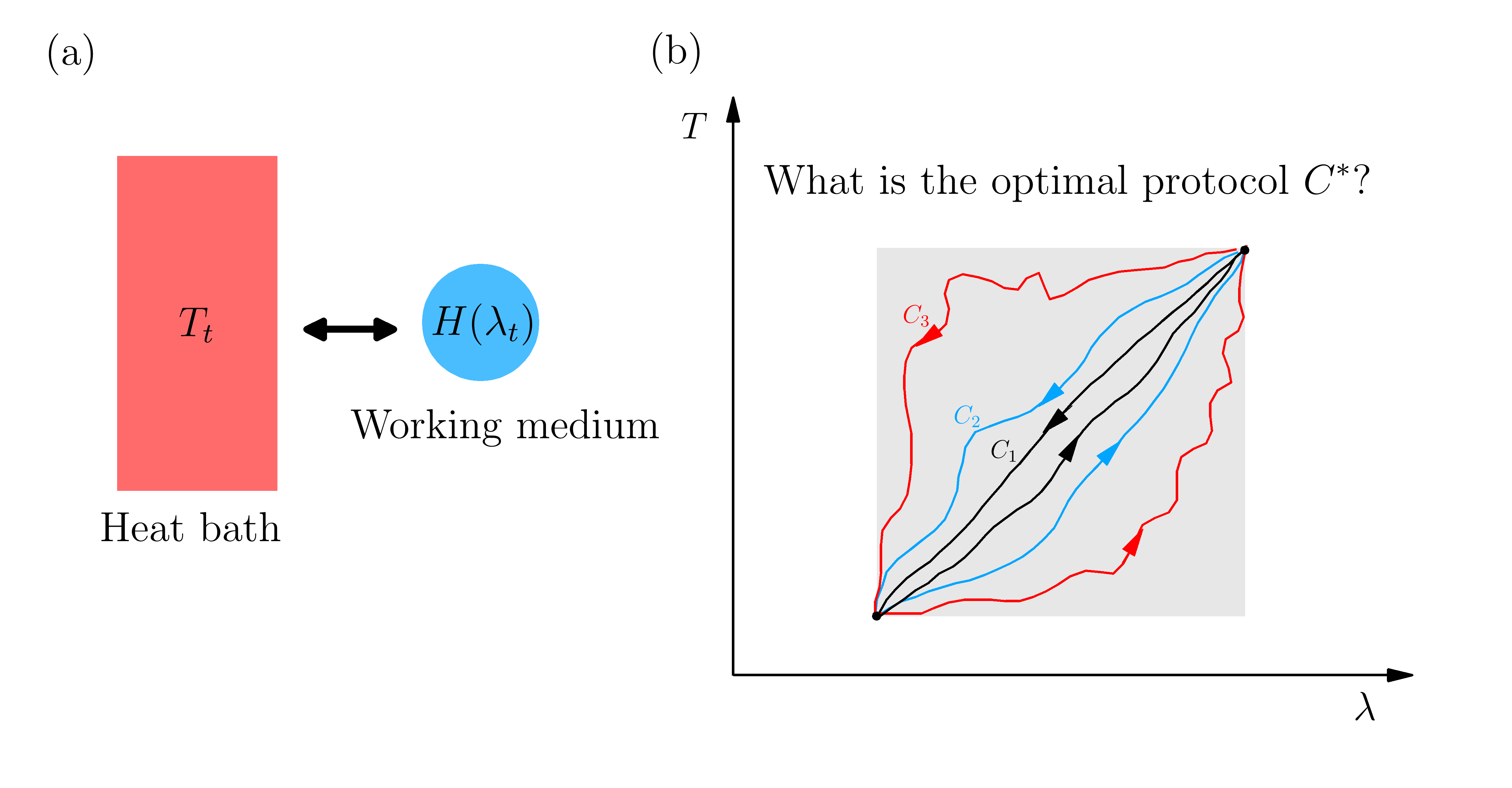}
\caption{\label{fig:1} Schematic Graphs. (a) The schematic graph of microscopic heat engine. It is a device that a working medium with controllable parameter $\lambda$ interacting with a heat bath with controllable temperature $T$. (b) The schematic graph of optimal cyclic protocol $C$ in the designated regime (gray). Arrows indicate the counterclockwise direction of $C$. The limitation in this paper is that two corners of the regime must on $C$ as shown in (b).}
\end{figure}

To apply the numerical method to the engine cycle, we need to fix the target function to optimize. In this paper, we consider the heat engine with slow driving parameters near quasi-static motion, where the adiabatic expansion from the instantaneous equilibrium state can formulate the thermodynamic quantities. We here introduce the theory in Ref.\cite{brandner2020thermodynamic} which addresses the relation between the power and thermodynamic efficiency in heat engines with slowly varying parameters. As shown in Fig. 1 (a), we consider a microscopic heat engine which consists of a working medium with the Hamiltonian $H(\lambda)$ and a heat bath with the temperature $T$. Here, $\lambda$ is a time-dependent control parameter for the working medium, and the temperature also changes continuously in time. Both $\lambda$ and $T$ vary continuously and return to the original values after the period $\tau$ as shown in Fig. 1 (b). The heat engine cycle forms a closed path $C$ in the ($\lambda, T$)-plane. Note that in Fig. 1 (b) we especially fix the two points in the ($\lambda, T$)-plane, which will be discussed later in our numerical scheme.

We consider the time-dependent Markovian master equation for the density matrix $\rho_t$:
\begin{equation}
    \partial_t \rho_{t} = \mathbb{L}_t \rho_{t} \, . \label{eq:1}
\end{equation}
For instance, in the classical Brownian heat engine, one can employ the Fokker-Planck operator \cite{fokker1914mittlere} for the operator $\mathbb{L}_t$. When we consider the quantum discrete system for the working medium, we can employ the Lindblad equation \cite{kossakowski1972quantum,lindblad1976generators,gorini1976completely}. Following the standard framework of stochastic thermodynamics, the average work done by the working medium to the outer environment is written as follows
\begin{align}
    & W = 
        \int_0^{\tau}dt f_t^{\lambda} \dot{\lambda} \, , \label{eq:2-1} \\
    & f_t^{\lambda} \equiv 
        -\mathrm{Tr}\left [\rho_t \partial_{\lambda} H(\lambda) \right] \, , \label{eq:2-2}
\end{align}
where $\tau$ is the period of the cycle. Throughout this paper, the Boltzmann constant is set to be unity. In addition, the effective heat uptake over the cycle is defined as
\begin{align}
    & U = 
        -\int_0^{\tau}dt f_t^{T} \dot{T} \, , \label{eq:3-1} \\
    & f_t^{T} \equiv 
        -\mathrm{Tr}\left [\rho_t \ln \rho_t \right] \, . \label{eq:3-2}     
\end{align}
The functions $f_t^{\lambda}$ and $f_t^{T}$ are thermodynamic forces, which respectively play roles of the instantaneous generalized pressure and entropy of the working system. 

When the heat engine cycle contains continuously varying temperatures, the conventional thermodynamic efficiency is not very convenient. Hence, instead, Ref.\cite{brandner2020thermodynamic} introduced a more general definition on efficiency. Using the non-negativity of the entropy production rate and cyclic property, one can show the inequality $U-W \geq 0$. From this property, new efficiency $\eta$ is defined as
\begin{eqnarray}
    \eta \equiv W/U \leq 1 \, . \label{eq:4}
\end{eqnarray}
One can notice that $\eta$ is $1$ for reversible heat engines. The standard Carnot efficiency can be reproduced once suitably normalized for the engines with two temperatures such as the Carnot cycle \cite{brandner2020thermodynamic}.

Near the quasi-static regime where the controllable parameters change very slowly, the periodic state $\rho_t$ can be expanded from the quasi-static limit $1/\tau \rightarrow 0$ as $\rho_t \approx \rho_{eq, t} + \delta \rho_t + \delta(1/\tau^2)$, where $\rho_{eq, t}$ is the instantaneous equilibrium distribution, i.e, $\mathbb{L}_t \rho_{eq, t} = 0$. The first order $\delta \rho_t$ is solved through the equation,
\begin{eqnarray}
    \partial_t \rho_{eq,t} = \mathbb{L}_t \delta \rho_t \, . \label{eq:5}
\end{eqnarray}

Substituting the expressions into the thermodynamic forces, one obtains the following linear response form,
\begin{eqnarray}
    \left(\begin{matrix} 
        f_t^{T} \\ f_t^{\lambda} 
    \end{matrix}\right) 
    = 
        \left(\begin{matrix} 
            f_{eq,t}^{T} \\ f_{eq,t}^{\lambda} 
        \end{matrix}\right) + 
        \mathcal{R} 
        \left(\begin{matrix} 
            \dot{T} \\ \dot{\lambda} 
        \end{matrix}\right) 
    = 
        \left(\begin{matrix} 
            f_{eq,t}^{T} \\ f_{eq,t}^{\lambda} 
        \end{matrix}\right) + 
        \left(\begin{matrix} 
            \mathcal{R}^{TT} & \mathcal{R}^{T\lambda} \\
            \mathcal{R}^{\lambda T} & \mathcal{R}^{\lambda \lambda}
        \end{matrix}\right) 
        \left(\begin{matrix} 
            \dot{T} \\ \dot{\lambda} 
        \end{matrix}\right) \, , \label{eq:6}
\end{eqnarray}
where $f_{eq,t}^{T}$ and $f_{eq,t}^{\lambda}$ are thermodynamic forces of instantaneous equilibrium state. The matrix $\mathcal{R}$ is the adiabatic response matrix, which stands for the response against the perturbation from the instantaneous equilibrium state. The expression of the matrix elements in $\mathcal{R}$ depends on the choice of dynamics $\mathbb{L}_t$. Using this adiabatic expansion, it is shown in Ref.\cite{brandner2020thermodynamic}, the power $P$ and efficiency $\eta$ has the following trade-off relation
\begin{equation}
    P \leq (1-\eta)\left(\mathcal{W}/\mathcal{L}  \right)^2 \, , \label{eq:7}
\end{equation}
where $\mathcal{W}$ is the quasi-static work for given cycle $C$, and the quantity $\mathcal{L}$ is the thermodynamic length defined as 
\begin{align}
    & \mathcal{L} \equiv 
        \oint_C \sqrt{-g^{\mu \nu} d\mu d\nu} \, , 
    ~~~\nu, \mu = T~{\rm or}~\lambda\, , \label{eq:8-1}\\
    & g^{\mu \nu} = 
        (\mathcal{R}^{\mu \nu} + \mathcal{R}^{\nu \mu})/2 \, . \label{eq:8-2}         
\end{align}
The metric $g^{\mu \nu}$ is originally defined in the expansion of the function $U-W =\int_0^{\tau} dt g^{\mu \nu} \dot{\mu} \dot{\nu}$. The thermodynamic length is thought to be a measure of the minimal dissipation along the fixed contour $C$. The upper bound of power is linearly decreasing as the efficiency increases and its decreasing speed is determined by the combination of $\mathcal{W}/\mathcal{L}$. The quantity $\mathcal{W}/\mathcal{L} $ is purely geometrical quantity dependent solely on the given contour $C$. 

\subsection{Target function and available regime}\label{sec2subsec2}
It is known that the relation (\ref{eq:7}) is a tight bound for the power \cite{brandner2020thermodynamic}. Once the cycle is determined, the upper bound can be obtained by changing the speed of protocol along the given contour $C$. It means that when $\mathcal{W}/\mathcal{L}$ is extremely large for a certain cycle, high efficiency and large power output of the heat engine can be obtained simultaneously using such cycle. Hence, in this paper, we consider the numerical method to find the optimal cyclic protocol $C^{*}$ in the ($\lambda, T$)-plane to maximize $\mathcal{W}/\mathcal{L}$ with several conditions, i.e.,
\begin{eqnarray}
    \max_{C^*} \mathcal{W}/\mathcal{L}
    ~~~~{\rm with~the~condition~ } 
    \mathcal{W} \ge {\mathcal W}_0 \, , \label{eq:9}
\end{eqnarray}
within the available parameter regime in the $\lambda-T$ plane. Concerning the available parameter regime, for simplicity, we fix the two points as shown in Fig. 1 (b), and the available parameter regime is assumed to be inside the square given determined by the two points (grey shaded regime). Although this setup is simple, there are infinite numbers of candidates as contours even in this limited regime, and hence the optimization is highly nontrivial. We propose a numerical method to solve this problem based on two models described in the subsequent sections.

\section{Models}\label{section: 3}
\subsection{Classical overdamped Brownian particle heat engine}\label{sec3subsec1}
As the first example, we consider the overdamped Brownian particle heat engine. The working system here is an one-dimensional classical overdamped Brownian particle in the time-dependent potential $V(x, \lambda_t)$, where the quadratic potential $V(x, \lambda_t) = \lambda_t x^2/2$ is applied. The temperature of the system is also continuously controlled in time. Then the dynamics of the distribution is described by the Fokker-Planck equation:
\begin{align}
    & \partial_t \rho_t(x) = 
        \mathbb{L}_t \rho_t(x)\, , \label{eq:10-1} \\
    & \mathbb{L}_t =  
        \lambda_t \partial_x x + T_t \partial^2_x \, , \label{eq:10-2}
\end{align}
where $\rho_t(x)$ is the distribution of the particle at the position $x$ and time $t$. The instantaneous equilibrium state of the distribution is given by
\begin{eqnarray}
    \rho_{eq,t}(x) = \sqrt{\omega/\pi}e^{-\omega x^2} \, , \label{eq:11}
\end{eqnarray}
where $\omega=\lambda/2T$. To solve Eq. (\ref{eq:10-1}) in the slow driving case near quasi-static motion, the distribution $\rho_t(x)$ is expanded up to the first order. The first order of the distribution is explicitly given as 
\begin{equation}
    \delta \rho_t (x) = \frac{\dot{\omega}}{4\omega\lambda} \rho_{eq} (2\omega x^2 - 1) \, . \label{eq:12}
\end{equation}
The derivation of this expression is explained in the Appendix A. Inserting the expression (\ref{eq:12}) to the thermodynamic forces (\ref{eq:3-1}) and (\ref{eq:3-2}), 
the quasi-static generalized pressure $f_{eq,t}^{\lambda}$ and the adiabatic response matrix $\mathcal{R}$ in (\ref{eq:6}) are given as follows
\begin{align}
    & \mathcal{R} = 
        \left(\begin{matrix} 
            \mathcal{R}^{TT} & \mathcal{R}^{T\lambda} \\
            \mathcal{R}^{\lambda T} & \mathcal{R}^{\lambda \lambda}
        \end{matrix}\right) = 
        \frac{1}{4\lambda^2} \left(\begin{matrix} 
            -T/\lambda & 1 \\
            1 & -\lambda/T 
        \end{matrix}\right) \, , \label{eq:13-1} \\
    & f_{eq,t}^{\lambda} = 
        -\frac{T}{2\lambda} \, . \label{eq:13-2}
\end{align}
From these expressions, one can arrive at the expression of the quasi-static work and thermodynamic length for the given cycle $C$ 
\begin{align}
    \mathcal{W} &= 
        -\oint_C \frac{T}{2\lambda} d\lambda \, ,  \label{eq:14-1}\\                
    \mathcal{L} & = 
        \oint_C \sqrt{\frac{T}{4 \lambda^3}} \left|d\lambda -\frac{\lambda}{T} dT\right| \, . \label{eq:14-2}
\end{align}

\subsection{Single-qubit heat engine}\label{sec3subsec2}
Another example we consider is the single-qubit heat engine. This model is also employed in Ref. \cite{brandner2020thermodynamic} in a different context.
The engine consists of a qubit surrounded by a thermal environment with controllable temperature $T$. The Hamiltonian of the qubit is
\begin{equation}
    H_{\lambda} =
        -\frac{\hbar \Omega}{2}
        (\varepsilon \sigma_x + \sqrt{\lambda^2-\varepsilon^2}\sigma_z)\, . \label{eq:15}
\end{equation}
Here, $\sigma_x$ and $\sigma_z$ are Pauli matrices, $\hbar\Omega$ denotes overall energy scale, setting to be unity. The term $\varepsilon$ causes the quantum coherence and hence, the dynamics become classical once we set $\varepsilon=0$. The parameter $\lambda$ is time-dependent. We assume $\varepsilon \geq 0$ and $\lambda \geq \varepsilon$. 

The equation of motion with respect to the density matrix $\rho_t$ of the qubit is described by the Lindblad equation
\begin{equation}
    \mathbb{L}_{t} X = 
        -\frac{i}{\hbar}[H_{\lambda}, X] 
        + \sum_{\sigma}
        \left(
            [V^{\sigma}X, V^{\sigma\dag}] 
             + [V^{\sigma}, X V^{\sigma\dag}]
        \right) \, , \label{eq:16}
\end{equation}
where $V^{\sigma}$ is the jump operator satisfying the quantum detailed balance condition, $[H_{\lambda}, V^{\pm}] = \pm \hbar\Omega\lambda V^{\pm}$, and ${\rm Tr}\left[V^{\pm}V^{\pm\dag}\right]=\pm\Gamma\Omega\lambda/\left[1-\exp{\left[\mp\hbar\Omega\lambda/T\right]}\right]$. From the adiabatic approximation, the quasi-static work $\mathcal{W}$ is given as \cite{brandner2020thermodynamic}
\begin{align}
    \mathcal{W} &= \frac{\hbar\Omega}{2}\oint_C \tanh{[\hbar\Omega\lambda/2T]} d\lambda \, , \label{eq:17}
\end{align}
We can calculate the thermodynamic length by Eq. (\ref{eq:8-1}). Especially for $\varepsilon =0$, the thermodynamic length is given by $\mathcal{L}_d$ \cite{brandner2020thermodynamic}:
\begin{align}
    \mathcal{L}_d &= 
        \oint_C \sqrt{-g_d^{\mu \nu} d\mu d\nu} = 
        \oint_C \sqrt{\frac{R}{4 \lambda T}} \left|d\lambda -\frac{\lambda}{T} dT\right| . \label{eq:18} 
\end{align}
When $\varepsilon \neq 0$, we obtain different expressions containing the quantum effect. See \cite{brandner2020thermodynamic} and Appendix B for the details.

It is interesting that $\mathcal{L}_d$ shows the similar expression to $\mathcal{L}$ of the classical Brownian heat engine (Eq. (\ref{eq:14-2})). When thermodynamic length takes its minimum value 0, the absolute terms in both Eq. (\ref{eq:14-2}) and Eq. (\ref{eq:18}) are zero and give the differential equation,
\begin{eqnarray}
    d\lambda = \frac{\lambda}{T}dT \, , \label{eq:20}
\end{eqnarray}
One can solve this equation directly and find that $\lambda$ is linearly proportional to $T$, i.e., $\lambda = \kappa T$, where $\kappa$ is a time-independent constant. In this situation, the corresponding quasi-static work is also zero. This extreme situation is avoided by imposing a constraint on the adiabatic work ${\cal W} \ge {\cal W}_0 > 0$ as in (\ref{eq:9}).

\section{Method}\label{section: 4}
\subsection{Parametrization}\label{sec4subsec1}
\begin{figure*}[htp]
    \includegraphics[width=1\textwidth]{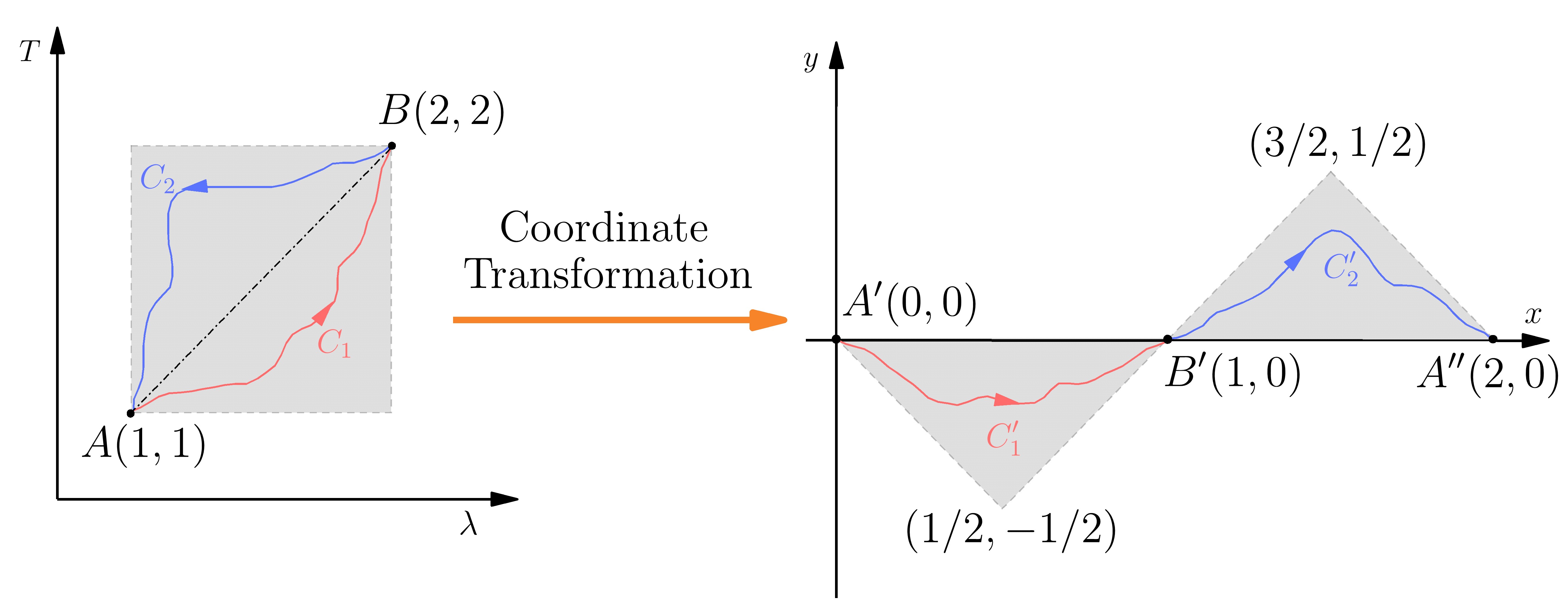}
    \caption{\label{fig:2} Schematic of the parametrization. We consider the cycle that passes two points A and B inside the grey regime. The original cycle on $T-\lambda$ plane is mapped onto the $x-y$ plane as in the right figure. The map is described by Eq. (\ref{eq:22}) and (\ref{eq:23})  }
\end{figure*}
The optimization problem (\ref{eq:9}) illustrated in section \ref{sec2subsec2} is actually a two-dimensional variational optimization problem that is difficult to solve analytically. Instead, we deal with it numerically.

We impose the following constraints for the cycle. (i) The cycle should be a simple closed curve, which means that the curve has no intersections. Otherwise, the protocol can be regarded as several sub-cycles. (ii) As explained in the section \ref{sec2subsec2}, we impose the constraint that the cycle must pass through the point $A$ and $B$, and we limit an available region to a square region made by the two points (see Fig. 1 (b) or the grey region in Fig. 2). (iii) We also impose the condition that the quasi-static work along the cycle must be larger than a given constant, i.e., $\mathcal{W} \geq {\mathcal W}_0$.

Concerning the constraint (ii), the available regime $D$ is mathematically described as
\begin{equation}
    D = \{(\lambda, T)|\lambda \in [1,2], T \in [1,2]\} \, . \label{eq:21}
\end{equation}
Because controllable parameters, such as the temperature $T$ and strength of potential or shifting energy $\lambda$, are usually finite. Note that even we limit the available regime as above, finding an optimized cycle is highly nontrivial.

Based on the above constraints on the cycle, we employ the following parametrizing method. See the schematic graph in Fig. 2. Assume that the optimal cycle $C^*$ must path through two corners of the available square regime $D$: $A(1,1)$ and $B(2,2)$ which divide the path into two parts, $C_1: A \rightarrow B$ and $C_2: B \rightarrow A$. We then map the closed curve $C^*$ into an unclosed curve $C'$ by the following coordinate transformations (See Fig. 2, again). We first translate coordinates so that $A$ can overlap with the origin, and rotate them $\pi/4$ counterclockwise to ensure that $B$ lies in the horizontal axis. Next, we fold the part $C_2$ into $C_2'$ centered on $B$. Thus, $A$ is folded to the other side of $B$ along the horizontal axis. After zooming, the intersects between the new curve $C'$ and the horizontal axis are $A'(0,0)$, $B'(1,0)$ and $A''(2,0)$. Then, $\lambda$ and $T$ can be expressed by the new coordinates $x$ and $y$, i.e.,
\begin{equation}
    \left\{
        \begin{array}{lcl}
            \lambda = x - y + 1\\
            T = x + y + 1 \\
        \end{array}
    \right.
    \text{for} \qquad C_1' \, , \label{eq:22}
\end{equation}
and
\begin{equation}
    \left\{
    \begin{array}{lcl}
        \lambda = 3 - x - y\\
        T = 3 - x + y \\
    \end{array}
    \right.
    \text{for} \qquad C_2' \, , \label{eq:23}
\end{equation}
where $C_1'$ and $C_2'$ respectively corresponds $C_1$ and $C_2$ in original curve. We can express them respectively by functions of $x$, $g_1(x)$ and $g_2(x)$. The total quasi-static work and thermodynamic length can then be expressed as the summations of the quasi-static works and thermodynamic lengths for $C_1$ and $C_2$ respectively. 
Thus, our optimization problem becomes finding the optimal $g_1(x)$ and $g_2(x)$ to maximize $\mathcal{W}/\mathcal{L}$, i.e.,
\begin{eqnarray}
    \max_{g_1,g_2} \frac{\mathcal{W}_1+\mathcal{W}_2}{\mathcal{L}_1+\mathcal{L}_2} \, , \label{eq:24}
\end{eqnarray}
where $\mathcal{W}_m$ and $\mathcal{L}_m$~($m=1,2$) are the quasi-static works and the thermodynamic lengths for the contour $C_m$, respectively.

In addition, we use the Fourier representation for the functions $g_m (x)~(m=1,2)$: 
\begin{eqnarray}
    g_m(x; \bm{b}^{(m)}) &= \sum_{j=1}^{n} b^{(m)}_j \sin (j \pi x) \, ,  ~ m = 1 \, , 2 \, . \label{eq:25}
\end{eqnarray}
Finding optimal path now reduces to finding the Fourier components, and hence the corresponding $\mathcal{W}_m$ and $\mathcal{L}_m$ is respectively parametrized by $\bm{b}^{(m)}$ as $\mathcal{W}_m(\bm{b}^{(m)})$ and $\mathcal{L}_m(\bm{b}^{(m)})$.

Finally, we write the constraints (i), (ii) and (iii) in terms of the functions $g_m(x; \bm{b}^{(m)})$. Note that without loss of generality, we choose the condition that $g_1(x) \leq 0$ and $g_2 \geq 0$ to satisfy the Constraint (i). Then, the expressions for the constraints can be written as follows
\begin{equation}\label{eq:26}
    \begin{array}{lcl}
        G_1(x;\bm{b}^{(1)}) = |g_1(x; \bm{b}^{(1)}) + \frac{x}{2}|     - \frac{x}{2} \leq 0 & & x \in [0,0.5)\\
        G_2(x;\bm{b}^{(1)}) = |g_1(x; \bm{b}^{(1)}) +
            \frac{1-x}{2}| - \frac{1-x}{2} \leq 0& & x \in [0.5,1)\\
        G_3(x;\bm{b}^{(2)}) = |g_2(x; \bm{b}^{(2)}) +
            \frac{1-x}{2}| - \frac{x-1}{2} \leq 0& & x \in [1,1.5)\\
        G_4(x;\bm{b}^{(2)}) = |g_2(x; \bm{b}^{(2)}) +
            \frac{x-2}{2}| - \frac{2-x}{2} \leq 0& & x \in [1.5,2)\\
        \mathcal{W}(\bm{b}^{(m)}) = \sum_{m=1}^2
            \mathcal{W}_m(\bm{b}^{(m)}) \geq \mathcal{W}_0 \, .
    \end{array}
\end{equation}

Consequently, our optimization problem (\ref{eq:9}) is reduced to finding the optimal parameters $\bm{b}^{(m)}$ under constraints (\ref{eq:26}) to maximize $\mathcal{W}/\mathcal{L}$, i.e., 
\begin{equation}\label{eq:27}
    \begin{aligned}
        \max_{\bm{b}^{(m)}} \frac{\mathcal{W}(\bm{b}^{(m)})}{\mathcal{L}(\bm{b}^{(m)})}, ~~~~{\rm with~the~condition~}(\ref{eq:26})
    \end{aligned}
\end{equation}

\subsection{Genetic Algorithm (GA)}\label{sec4subsec2}
Studies of computer sciences provided many optimization algorithms to solve the problem above. In this paper, we use the Genetic Algorithm (GA) with penalty functions. GA is an optimization method via simulating the natural selection \cite{whitley1994genetic}, which is originally inspired by the process of natural selection and biological evolution based on genetic mechanism. Focusing on all individuals in a population, GA randomly searches the optimal solution in coded parameter space. This gives its well global searching ability. To establish GA on our problem, the following five parts are needed to be carefully considered, including parameter-coding, initialization, design of fitness function and genetic operations (selection, crossover, and mutation), and settings of control parameters.

First, we consider the parameter-coding process. For our problem, the parameter vector to be optimized is $\bm{b}^{(m)}$. Thus, one can uniquely specify some imaginary creature by the values of $\bm{b}^{(m)}$. Similar to the gene sequence in the real world, if one regards $\bm{b}^{(m)}$ as the unique character of the imaginary creature, its binary notation vector $\hat{\bm{b}}^{(m)}$ then can be regarded as the corresponding unique gene sequence. The length of the binary notation can be adjusted by a factitiously determined precision $\Delta$. 

Second, in order to simulate the natural selection process, a population which consists of $N$ arbitrary imaginary creature is needed. Initially, it is a set of randomly-selected parameter vectors, i.e., $G = \{\hat{\bm{b}}_1^{(m)},\hat{\bm{b}}_2^{(m)}, ..., \hat{\bm{b}}_N^{(m)}\}$. 

Third, natural selection process obsoletes the creature according to its fitness, which is described by the fitness function in GA. Basically, the fitness function is often set to be the target function. For our problem, considering the constraints (\ref{eq:26}), we transform the problem (\ref{eq:27}) into the following one without constraints with the help of penalty function method,
\begin{equation}\label{eq:28}
    \begin{aligned}
        \min_{\bm{b}^{(m)}} \quad & F(\bm{b}^{(m)}) = -\frac{\mathcal{W}(\bm{b}^{(m)})}{\mathcal{L}(\bm{b}^{(m)})} + P(\bm{b}^{(m)}). & \\
    \end{aligned}
\end{equation}
Here, $P(\bm{b}^{(m)})$ is the penalty function,
\begin{eqnarray}\label{eq:29}
    P(x;\bm{b}^{(m)}) &&= 
    \sigma \sum_{k=1}^{n_c} \sum_{i=1}^{4} \left|\max(0, G_i(x_k;\bm{b}^{(m)})\right| \nonumber \\
    &&+ \sigma \left|\max(0, \mathcal{W}(\bm{b}^{(m)})\right|.
\end{eqnarray}
by equally dividing the range $[0, 2]$ of $x$ with $n_c$ points. $\sigma$ is a large penal constant. Hence, the fitness function in our problem is set to be $F(\bm{b}^{(m)})$. In the selection step, we calculate $F(\bm{b}^{(m)})$ for every creature in the population $G$. If some creature has a low value of fitness, it may die out with a larger probability than others. We select $N_s$ creatures in $G$ to survive. Please see Appendix D for details.

Moreover, in order to adapt to the environment, these $N_s$ surviving creatures must reproduce their children. Parents are equally randomly selected in the surviving creatures and a pair of parents can only reproduce one child. By gene mutation and gene crossover process, genes in children can be optimized. In situations considered in this paper, two random sites of each dimension of the binary notation vector $\hat{\bm{b}}^{(m)}$ are selected to be the crossover points on both parents' gene sequences, and each fragment divided by the selected sites has a probability to be inherited by their child. In the gene mutation process, random sites of each dimension of the binary notation vector $\hat{\bm{b}}_1^{(m)}$ has a probability to change to its opposite value ($0$ or $1$) only. Please see Appendix E for details.

Thus, the remaining $N_s$ parents and the new $N-N_s$ child make up the new population. After enough such iterations, the process of natural selection ensures that the fitness of each individual in the population can be optimized.

\subsection{Example of optimization without constraints}\label{sec4subsec3}

\begin{figure}[htp]
\includegraphics[width=\textwidth]{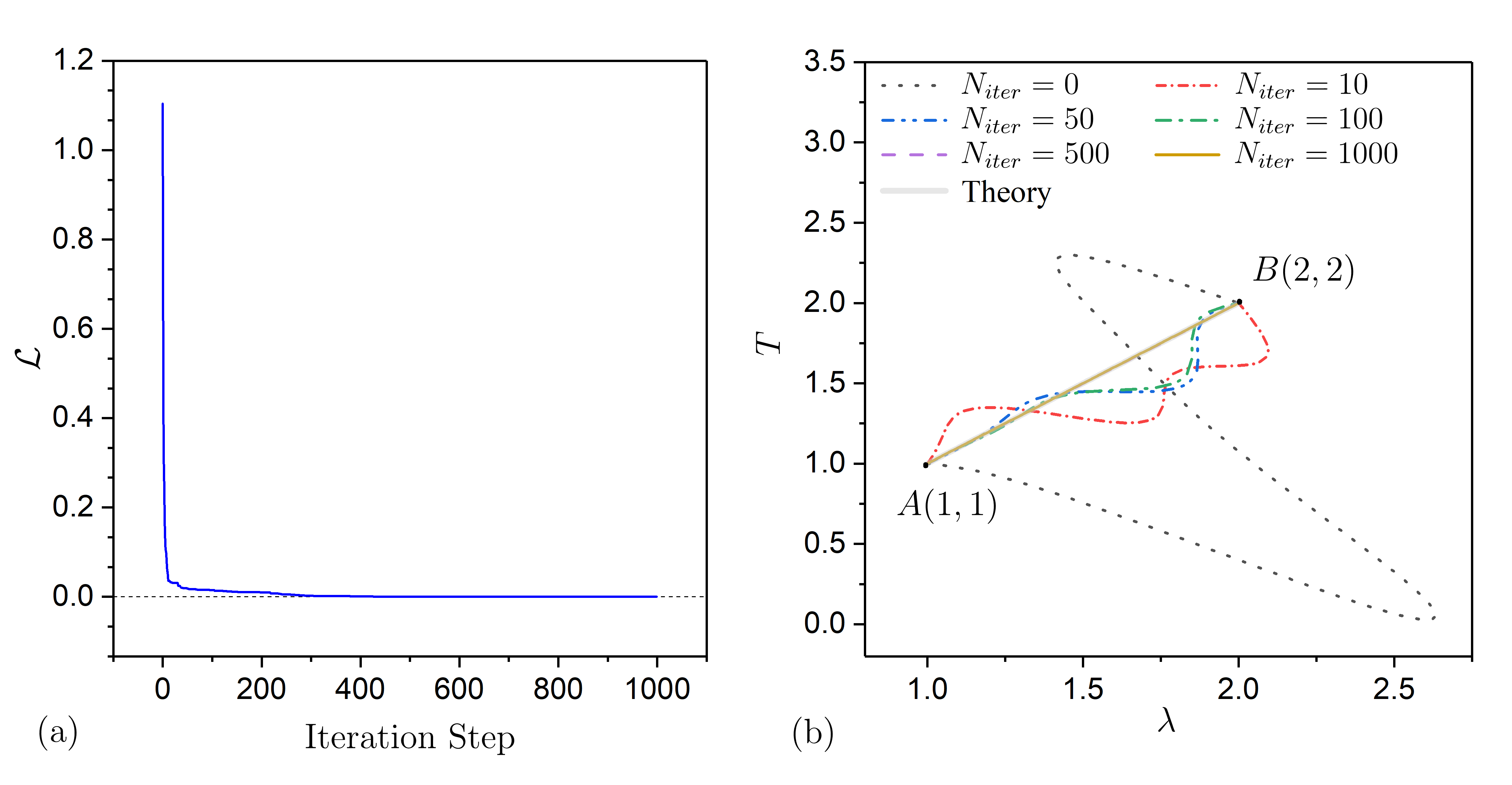}
\caption{\label{fig:3} Numerical results of optimal protocol that minimize the thermodynamic length of classical overdamped Brownian particle heat engine. (a) Relationship Between Objective Function $\mathcal{L}$ and Iteration Step of Genetic Algorithm. They show that $\mathcal{L}$ decreases and converge to 0 as iteration step increases from $0$ to $1000$. (b) Optimal protocols for different iteration steps. The gray line is the theoretical result $T = \lambda$. We set the Fourier bases to be $5$ here.}
\end{figure}

To demonstrate that the GA can correctly find the optimal solution for our problem, we present an analytically solvable problem in this section. The problem is to find the optimal protocol $\gamma^*$ in the parameter plane $\lambda-T$ from $A(1,1)$ to $B(2,2)$ that minimizes the minimum thermodynamic length $\mathcal{L}$. We consider this problem for the classical overdamped Brownian particle, i.e., 
\begin{eqnarray}\label{eq:30}
    \min_{\gamma : A \to B} \mathcal{L}.
\end{eqnarray}
As discussed in Section \ref{sec2subsec2}, we can theoretically find the optimal path $\gamma^*$. The solution is a simple line $T=\lambda$, where the corresponding minimum thermodynamic length $\mathcal{L}^* = 0$. Numerically, we deal with this problem by employing the parametrization method and the GA above. Instead of considering the whole cycle, we only need to parametrize half of it (i.e., $C_1$) like the form of Eq. (\ref{eq:26}). Besides, we do not have any constraints on $\gamma$ here. Results are shown in Fig. 3. One can notice that as the iteration steps increases, the protocol gradually converges to the theoretical solution (Fig. 3 (a)) and the target function $\mathcal{L}$ converges to the theoretical prediction $0$ (Fig.3 (b)). It indicates that our method with the GA is reliable.

\section{Application to heat engines}\label{section: 5}

In this section, we present a numerical demonstration obtained by the above method, using the target function $F$ in (\ref{eq:28}). We show that the numerical recipe works well for optimization problems in heat engines. We maximize the ratio $\mathcal{W}/\mathcal{L}$ for the models that we have introduced in section \ref{section: 3}, i.e., the classical Brownian heat engine, the classical and quantum single-qubit heat engines. As shown below, large ${\cal W}$ requires a large area of the cycle, while small ${\cal L}$ needs a small area. Hence, there is a sort of trade-off to maximize the combination $\mathcal{W}/\mathcal{L}$. This competition leads to several nontrivial behaviors on the heat engine cycle in the available squared region depicted in Fig.1 (b).

\begin{figure}[htp]
\includegraphics[width=\textwidth]{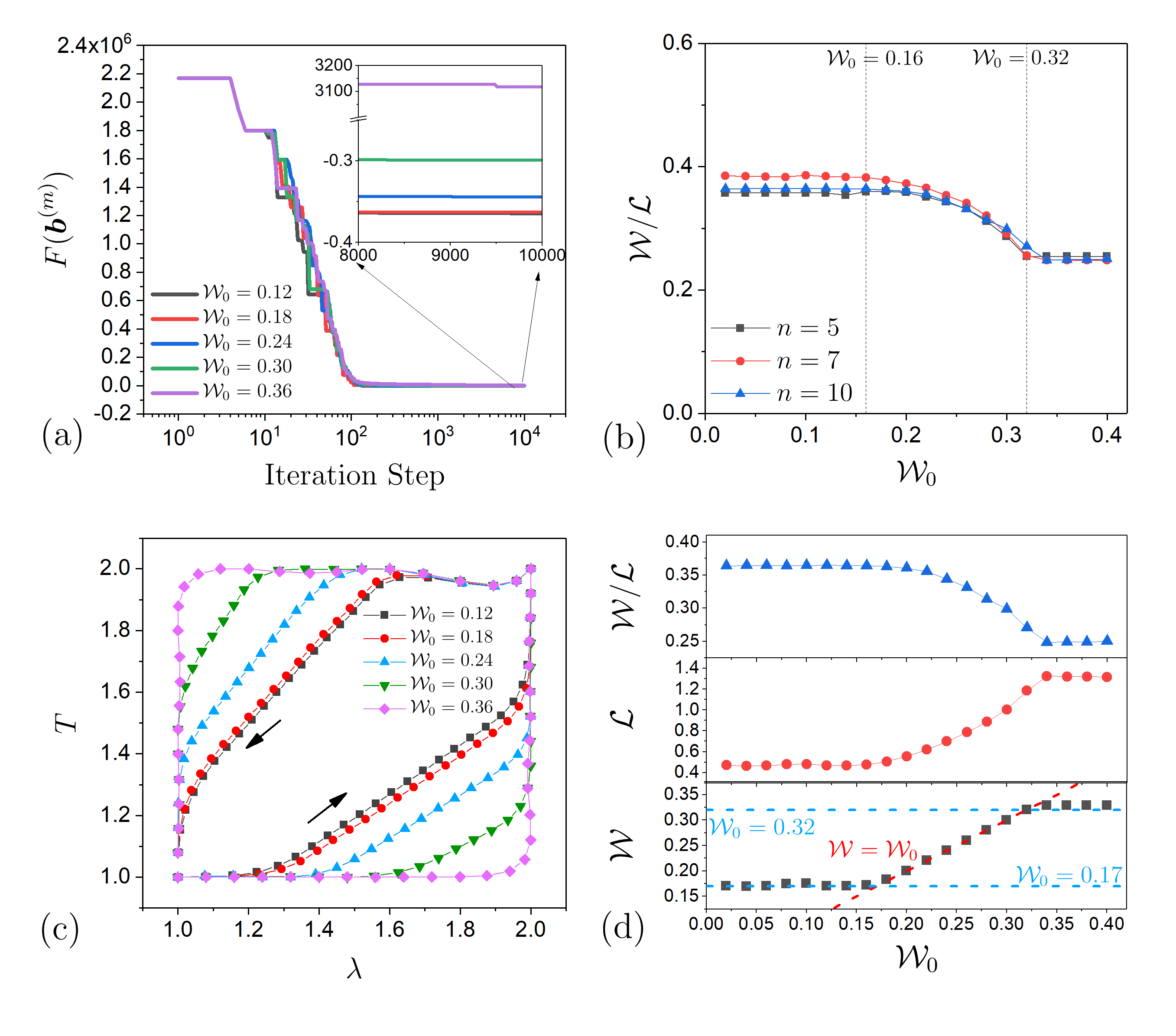}
\caption{\label{fig:4} Numerical results of optimal thermodynamic cycle for classical overdamped Brownian particle heat engine. (a) shows that objective function $F(\bm{b}^{(m)})$ decreases and converge to a constant as iteration step increases from $1$ to $10^4$. Lines of different colors denote different values of lower bound $\mathcal{W}_0$ of quasi-static work. Details of this graph is shown in the subgraph. (b) indicates the effect of the number of Fourier Bases on $\mathcal{W}/\mathcal{L}$. (c) shows that how optimal cycles behave in the parameter plane $T-\lambda$. Colors of these lines corresponds to those of Left columns. Arrows stand for that the directions of cycles are counterclockwise along which the quasi-static work is positive. (d) is the relation of optimal quantities versus $\mathcal{W}_0$. Quasi-static work $\mathcal{W}$, thermodynamic length $\mathcal{L}$ and the ratio $\mathcal{W}/\mathcal{L}$ is presented by the blue, red and black line, respectively. }
\end{figure}

\subsection{Brownian heat engine}

We first show the results for the Brownian heat engine. In this case, we set a finite constraint on the quasi-static work, i.e.,  ${\cal W} \ge {\cal W}_0 ~(>0)$. Fig. \ref{fig:4} (a) shows a numerical saturation as a function of iteration steps. One can notice that the target function $F$ converges to some constant after sufficient large iteration steps for each situation. Here, we use the same random seed in the numerical calculation as an initial state, and hence, different curves overlap for different $\mathcal{W}_0$ at small iteration steps in Fig. \ref{fig:4} (a). In Figs.\ref{fig:4} (a), (c) and (d), we use the Fourier modes $n=10$, which is enough for accuracy, as indicated in Fig. \ref{fig:4} (b).

Once target functions converge, optimal thermodynamic cycles can be obtained, which are shown in Fig. \ref{fig:4} (c). We also show the saturated values of ${\cal W}/{\cal L}$, ${\cal L}$, and ${\cal W}$ as a function of constraint ${\cal W}_0$ in Fig. \ref{fig:4} (d). We discuss these results. As shown in Fig. \ref{fig:4} (c), by imposing the constraint on the quasi-static work $\mathcal{W}$, one can see that as increasing the constraint ${\cal W}_0$, the optimal cycle approaches to the boundary of $D$ which is the allowable maximum cycle in the domain. This supports the intuition that a large work ${\cal W}$ requires a large area.

The work $\mathcal{W}$ reaches its maximum value ($\sim 0.32$) shown in Fig. \ref{fig:4} (d). The available domain $D$ plays the role of an upper bound on the quasi-static work. To produce the maximum work and maximum $\mathcal{W}/\mathcal{L}$ simultaneously, the heat engine is required to be operated inside the boundary of $D$. Particularly, if $\mathcal{W}_0$ exceeds the upper limit ($\sim 0.32$), i.e., the constraint $\mathcal{W} \geq \mathcal{W}_0$ is no longer applicable. In this case, the term $\sigma \left|\max(0,\mathcal{W}(\bm{b}^{(m)})\right|$ in the penalty function (\ref{eq:29}) can never be avoided. Our algorithm can still minimize the target function $F$ and reduce the penal term above as much as possible. The consequence of this trend is that the quasi-static work can still be maximized to the upper limit, but the corresponding optimal cycle will no longer change. This is why the cycles overlap with each other when $\mathcal{W}_0 > 0.32$ in Fig. \ref{fig:4} (c).

By imposing a large constraint ${\cal W}_0$ as above, one can see that a large area in the cycle is generally necessary for large adiabatic work ${\cal W}$. On the other hand, we also need to note that a small area in the cycle is required for a small thermodynamic length ${\cal L}$. Hence, in general, there is a sort of trade-off between obtaining large ${\cal W}$ and reducing ${\cal L}$. As decreasing the constraint ${\cal W}_0$, the effects arising from this trade-off become significant. Let us discuss the regime of ${\cal W}_0 < 0.32$ in the bottom figure of Fig. \ref{fig:4} (d). The figure shows that in this regime, the quasi-static work along the optimal cycle tends to equal $\mathcal{W}_0$. It is also equivalent to that our algorithm tries to obtain a small area of the engine cycle, which leads to small values of ${\cal L}$. Hence, the numerical results non-trivially show that to obtain the maximizing the rate ${\cal W}/{\cal L}$, obtaining small adiabatic work $\mathcal{W}$ is a good strategy. This also implies that the constraint on $\mathcal{W}$ limits the maximum value of $\mathcal{W}/\mathcal{L}$ in this regime. We also notice that the corresponding optimal cycle also gradually shrinks from the boundary (the cases ${\cal W}_0 =0.18, 0.24$ and $0.30$) in Fig. \ref{fig:4} (c). We believe that this behavior appears as a result of nontrivial trade-off properties to ${\cal W}/{\cal L}$.

Until $\mathcal{W}_0 \leq 0.17$, flat behavior appear again in the graph of $\mathcal{W}$, $\mathcal{L}$ and $\mathcal{W}/\mathcal{L}$. These platforms indicate that there will be no optimal cycle better than that of $\mathcal{W} = 0.17$, i.e., the cycle we obtain here is the global optimal cycle that can maximize $\mathcal{W}/\mathcal{L}$ in domain $D$.

We now recall that the value $\mathcal{W}/\mathcal{L}$ determines whether the large power and large efficiency can be obtained simultaneously, as shown in the general theory (\ref{eq:7}). As shown in Fig. \ref{fig:4} (d), the maximum value of $\mathcal{W}/\mathcal{L}$ is about $0.40$, which is finite, and hence large power and large efficiency cannot be obtained simultaneously in this system. However, we stress that based on our approach, one can systematically find the optimum cycle for the Brownian heat engine.

\begin{figure}[htp]
\includegraphics[width=\textwidth]{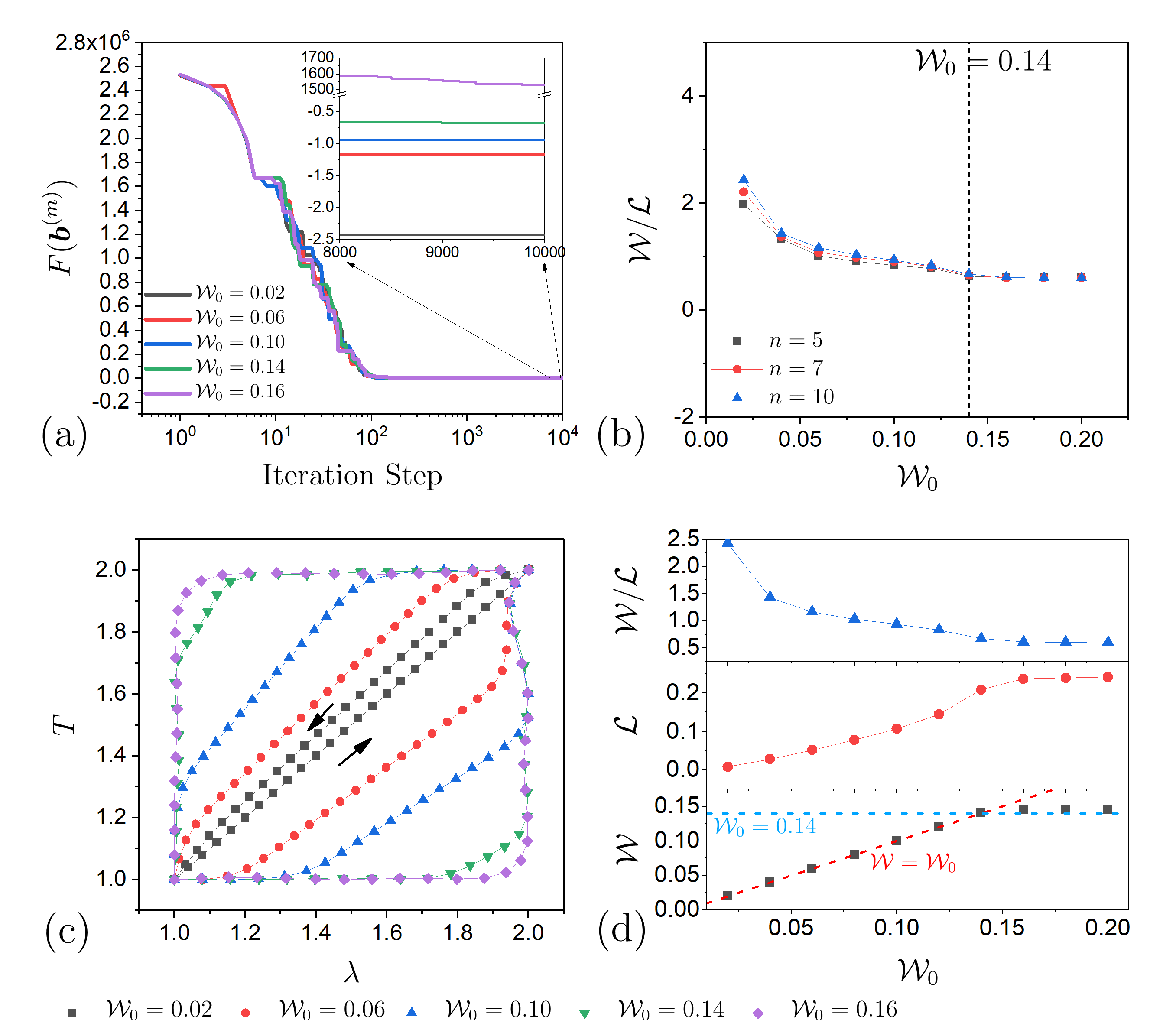}
\caption{\label{fig:5} Numerical results of optimal thermodynamic cycle for quantum single-qubit heat engine in quasi-classical limit ($\epsilon=0$). (a) shows that objective function $F(\bm{b}^{(m)})$ decreases and converge to a constant as iteration step increases from $1$ to $10^4$. Lines of different colors denote different values of lower bound $\mathcal{W}_0$ of quasi-static work. Details of this graph is shown in the subgraph. (b) indicates the effect of the number of Fourier Bases on $\mathcal{W}/\mathcal{L}$. (c) shows that how optimal cycles behave in the parameter plane $T-\lambda$. Colors of these lines corresponds to those of Left columns. Arrows stand for that the directions of cycles are counterclockwise along which the quasi-static work is positive. (d) is the relation of optimal quantities versus $\mathcal{W}_0$. Quasi-static work $\mathcal{W}$, thermodynamic length $\mathcal{L}$ and the ratio $\mathcal{W}/\mathcal{L}$ is presented by the blue, red and black line, respectively. }
\end{figure}

\subsection{Single-qubit heat engine}
We next show that our numerical scheme also works well in the single-qubit heat engines. In this case, we can consider the effect of the quantum coherence $\varepsilon$ in the dynamics. Hence, we consider two cases, i.e., the classical case and the quantum case. In the classical case, we set the quantum coherence $\varepsilon$ to zero in the Hamiltonian (\ref{eq:15}), and we set a finite constraint on the adiabatic work. In the quantum case, we set a finite value for the quantum coherence $\varepsilon$, and we do not impose a constraint on the adiabatic work, i.e., ${\cal W}_0 =0$.

\begin{figure}[htp]
\includegraphics[width=\textwidth]{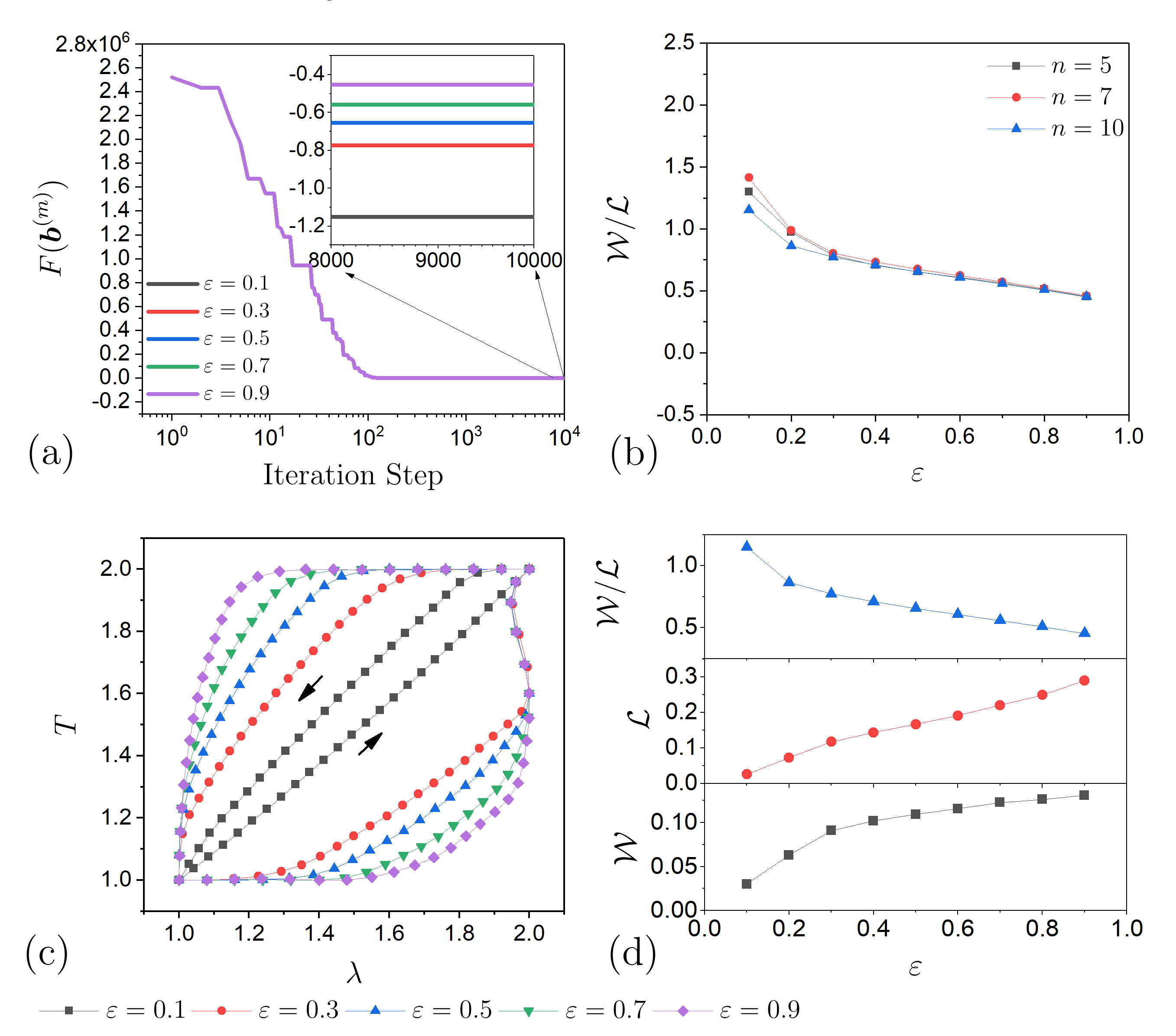}
\caption{\label{fig:6} Numerical Results of Optimal Cyclic Protocols for Quantum Single-qubit Heat Engine in Quantum Case. (a) shows that objective function $F(\bm{b}^{(m)})$ decreases and converge to a constant as iteration step increases from $1$ to $10^4$. Lines of different colors denote different values of quantum coherence factor $\epsilon$ of quasi-static work. Details of this graph is shown in the subgraph. (b) indicates the effect of the number of Fourier Bases on $\mathcal{W}/\mathcal{L}$. (c) shows that how optimal cycles behave in the parameter plane $T-\lambda$. Colors of these lines corresponds to those of Left columns. Arrows stand for that the directions of cycles are counterclockwise along which the quasi-static work is positive. (d) is the relation of optimal quantities versus $\epsilon$. Quasi-static work $\mathcal{W}$, thermodynamic length $\mathcal{L}$ and the ratio $\mathcal{W}/\mathcal{L}$ is presented by the blue, red and black line, respectively. }
\end{figure}

We first discuss the classical case $\varepsilon=0$. The results are shown in Fig. \ref{fig:5}. We use the Fourier mode $n=10$, which does not cause significant errors, as can be seen in Fig. \ref{fig:5} (b). This indicates that our numerical scheme works well even in discrete level heat engines. Regarding the behavior of cycles, a similar conclusion to the Brownian heat engine can be given.
Our numerical calculation shows that the value of $\mathcal{W}/\mathcal{L}$ is always finite, which implies that the large power and large efficiency cannot be obtained simultaneously, similarly to the Brownian heat engine case. 

We next look at the quantum case, i.e., $\varepsilon \neq 0$. We can see that our method works well as shown in Figs. \ref{fig:6}(a) and (b). As indicated in Fig. \ref{fig:6}, the quantum coherence factor $\varepsilon$ automatically provides a {\it constraint} on $\mathcal{W}$ when maximizing $\mathcal{W}/\mathcal{L}$, which reduces the performance of the heat engine. This is consistent with Brander and Saito's conclusion for the example cycle they introduced \cite{brandner2020thermodynamic}. For the available $\lambda$-$T$ plane, our numerical algorithm clarifies the best heat engine cycle even in the quantum case.

\section{Summary}\label{section: 6}

In summary, we propose a numerical optimization method based on genetic algorithm and penalty method to find the optimal thermodynamic cycle under the framework of adiabatic response regime. Two models driven by slow temperature and slow external forces in a designated domain are considered. The optimal thermodynamic cycle for both models under a classical, quasi-classical, and quantum case are successfully obtained by our method. The shrinking behaviors with the constraint on quasi-static work of these optimal cycles confirm that large power and high efficiency cannot coexist in our examples. Moreover, the quantum coherence factor is found to play a similar role to the constraint in quantum systems. Thanks to this property, the adiabatic response method can be applied on quantum systems, while there is an invalid situation where the system is always staying in the same instantaneous equilibrium state for the classical case. Our numerical method can be applied to other optimal-protocol-seeking problems as well. One of the relevant perspectives is trying to find the power-efficiency trade-off relations and optimal thermodynamic cycles in non-Markovian microscopic heat engines.

\begin{acknowledgements}
We thank K. Saito for useful discussion and encouragement.
\end{acknowledgements}

%
%

\bibliographystyle{spphys}       
\bibliography{references.bib}   

%
%

\appendices
\section*{Appendix A: Adiabatic Response Theory on Classical Overdamped Brownian Particle Heat Engine}\label{section: appendixa}
We introduce details on the classical overdamped Brownian particle heat engine discussed in Section \ref{sec3subsec1} of the main text. Specifically, we show how its finite-time thermodynamics can be analyzed under the adiabatic response regime.

As illustrated in the main text, the probability distribution $\rho(x, t)$ to find the Brownian particle at position $x$ at time $t$ is governed by the corresponding Fokker-Planck equation (\ref{eq:10-1}).

Our goal is to find the first two orders of adiabatic expansion on the generalized pressure $f_t^{\lambda}$ and the entropy $f_t^T$, which are defined in Eq. (\ref{eq:2-2}) and (\ref{eq:3-2}). Consider the quasi-adiabatic expansion of $\rho(x, t)$ (by the order of speed $1/\tau$),
\begin{equation}\label{eq:a1}
\rho(x, t) = \rho_{eq} + \delta \rho + o(1/\tau^2) \approx \rho_{eq} + \delta \rho
\end{equation}

Supposing that we denote $\omega = \lambda/2T$, we obtain $\rho_{eq}$, the instantaneous equilibrium state,
\begin{equation}\label{eq:a2}
\rho_{eq}(x,t) = \sqrt{\frac{\omega}{\pi}} e^{-\omega x^2}.
\end{equation}
Thus, we have this relationship by the normalization property of $\rho$,
\begin{equation}\label{eq:a3}
\int_{-\infty}^{+\infty} dx \delta \rho = 0
\end{equation}

Inserting Eq. (\ref{eq:a1}) into Eq. (\ref{eq:10-1}) and eliminating the high-order small term, we obtain
\begin{equation}\label{eq:a4}
\frac{\partial \rho_{eq}}{\partial t} = \mathbb{L}_t \delta \rho, 
\end{equation}

To solve Eq. (\ref{eq:a4}), we do the following Fourier transformation on both sides,
\begin{equation}\label{eq:a5}
\widetilde{\delta \rho}(k,t) = \mathcal{F}[\delta \rho(x,t)] = \frac{1}{\sqrt{2 \pi}} \int_{-\infty}^{+\infty} dx e^{-ikx} \delta \rho(x,t).
\end{equation}
Then, Eq. (\ref{eq:a4}) becomes,
\begin{equation}\label{eq:a6}
\mathcal{F}\left[ \frac{\partial \rho_{eq}}{\partial t} \right] =  -\lambda(t) k \frac{\partial \widetilde{\delta \rho}(k,t) }{\partial k} - T(t) k^2 \widetilde{\delta \rho}(k,t),
\end{equation}
where
\begin{eqnarray}\label{eq:a7}
\mathcal{F}\left[ \frac{\partial \rho_{eq}}{\partial t} \right] &&= \mathcal{F}\left[ \dot{\omega} \sqrt{\frac{\omega}{\pi}} e^{-\omega x^2} \left(\frac{1}{2 \omega} - x^2 \right)\right] \nonumber\\
&& = \frac{ \dot{\omega} }{4 \sqrt{2 \pi } \omega^2} k^2 e^{-\frac{k^2}{4 \omega}}
\end{eqnarray}

Thus, we obtain a differential equation of first-order of $k$ of $\delta \rho(k, t)$,
\begin{equation}\label{eq:a8}
\frac{\partial \widetilde{\delta \rho}(k,t)}{\partial k} + \frac{k}{2w} \widetilde{\delta \rho}(k,t) + \frac{ \dot{\omega} }{4 \sqrt{2 \pi } \omega^2 \lambda} k e^{-\frac{k^2}{4 \omega}} = 0.
\end{equation}

We can directly solve it,
\begin{equation}\label{eq:a9}
\begin{aligned}
\widetilde{\delta \rho}(k,t) &= C e^{-\int dk \frac{k}{2\omega}} \\
& \quad -e^{-\int dk \frac{k}{2\omega}} \int dk \frac{ \dot{\omega} }{4 \sqrt{2 \pi } \omega^2 \lambda} k e^{-\frac{k^2}{4 \omega}} e^{\int dk \frac{k}{2\omega}} \\
&= C e^{-\frac{k^2}{4 \omega}} - \frac{ \dot{\omega} }{8 \sqrt{2 \pi } \omega^2 \lambda} k^2 e^{-\frac{k^2}{4 \omega}}.
\end{aligned}
\end{equation}
By using inverse Fourier Transform on $\widetilde{\delta \rho}(k,t)$, we obtain,
\begin{eqnarray}\label{eq:a10}
\delta \rho(x, t) &&= \mathcal{F}^{-1} [\widetilde{\delta \rho}(k,t) ] \\
&& = C \sqrt{2\omega} e^{-\omega x^2} + \frac{\dot{\omega}}{4\omega\lambda} \sqrt{\frac{\omega}{\pi}} e^{-\omega x^2} (2\omega x^2 - 1) \nonumber
\end{eqnarray}

Using property (\ref{eq:a3}) of $\delta \rho(x, t)$, we can determine $C=0$. Therefore, first-order of adiabatic expansion of $\rho(x,t)$ is,
\begin{equation}\label{eq:a11}
\delta \rho(x, t) = \frac{\dot{\omega}}{4\omega\lambda} \rho_{eq} (2\omega x^2 - 1)
\end{equation}

Then, the generalized pressure $f_t^{\lambda}$ and the entropy $f_t^T$ can be calculated. The equilibrium part is,
\begin{equation}\label{eq:a12}
\begin{aligned}
f_{eq,t}^{\lambda} & = \langle  -\frac{\partial H}{\partial \lambda} \rangle = -\langle  x^2 \rangle/2 = -\frac{T}{2\lambda} \\ 
f_{eq,t}^{T} &= -\langle  \ln \rho(x,t) \rangle = \frac{1}{2} \left(1 + \ln \frac{2 \pi T}{\lambda}\right)
\end{aligned}
\end{equation}

Therefore, we have
\begin{equation}\label{eq:a13}
\begin{aligned}
f_t^{\lambda} & = \langle  -\frac{\partial H}{\partial \lambda} \rangle \approx -\frac{1}{2} \int dx (\rho_{eq} + \delta \rho) x^2\\
& = f_{eq, t}^{\lambda} - \frac{T}{4\lambda^3} \dot{\lambda} + \frac{1}{4\lambda^2} \dot{T} \\
\end{aligned}
\end{equation}

\begin{equation}\label{eq:a14}
\begin{aligned}
f_t^{T} &= -\langle  \ln \rho \rangle
\approx \int dx (\rho_{eq} + \delta \rho) \ln \rho_{eq} \\
& = f_{eq, t}^{T} + \frac{1}{4\lambda^2} \dot{\lambda} - \frac{1}{4\lambda T} \dot{T}.
\end{aligned}
\end{equation}

Thus, the adiabatic response matrix can be obtained,
\begin{equation}\label{eq:a15}
\mathcal{R}= {\left(\begin{array}{ccc}
\mathcal{R}^{\lambda\lambda} & \mathcal{R}^{\lambda T} \\
\mathcal{R}^{T \lambda} & \mathcal{R}^{TT}
\end{array} \right)} = 
\frac{1}{4\lambda^2}{
\left( \begin{array}{ccc}
-T/\lambda & 1 \\
1 & -\lambda/T 
\end{array} 
\right )}
\end{equation}

Then, the corresponding adiabatic work $\mathcal{W}$ is expressed as
\begin{equation}\label{eq:a16}
\mathcal{W} = \oint_c d\lambda f_{eq,t}^{\lambda} = -\int_0^{\Delta t}dt \frac{T}{2\lambda} \dot{\lambda}
\end{equation}
and the thermodynamic length $\mathcal{L}$ is denoted by
\begin{equation}\label{eq:a17}
\begin{aligned}
\mathcal{L} &= \oint_C \sqrt{-[\mathcal{R}^{\lambda \lambda} d\lambda d\lambda + (\mathcal{R}^{\lambda T} + \mathcal{R}^{ T \lambda}) d\lambda dT + \mathcal{R}^{TT} dTdT] } \\
&=\oint_C \sqrt{\frac{T}{4 \lambda^3}} \left|d\lambda -\frac{\lambda}{T} dT\right| \\
\end{aligned}
\end{equation}

\section*{Appendix B: Adiabatic Response Theory on Single-qubit Heat Engine}
In this section, we derive the expressions of thermodynamic length and quasi-static work of single-qubit heat engine under the framework of adiabatic response theory, which has been discussed in the Ref. \cite{brandner2020thermodynamic}

As illustrated before, the time evolution of the density matrix $\rho_t$ of the working medium is described by the Lindblad equation,
\begin{equation}
    \partial_t \rho_t = \mathbb{L}_t \rho_t \, . \label{eq:b1}
\end{equation}
Here $\mathbb{L}_t$ is the Lindblad generator and it is composed of two parts: the unitary evolution term of the bare system and the influence of the environment, i.e.,
\begin{equation}
    \mathbb{L}_t X \equiv (-i\mathrm{H}_t + \mathrm{D}_t)X \, , \label{eq:b2}
\end{equation}
where 
\begin{align}
    &\mathrm{H}_t X \equiv [H, X]/\hbar ~~~~ \rm{and} \nonumber \\
    &\mathrm{D}_t X \equiv \sum_{\sigma}
        \left(
            [V^{\sigma}X, V^{\sigma \dag}]+
            [V^{\sigma}, XV^{\sigma \dag}]
        \right) \, . \label{eq:b3}
\end{align}
According to the result of Ref. \cite{brandner2020thermodynamic}, 
the state $\rho_t$ can be determined by an adiabatic perturbation theory. By inserting the ansatz 
\begin{equation}
    \rho_t = \rho_{eq,t} + \delta \rho_t \label{eq:b4}
\end{equation}
into (\ref{eq:b3}), after simplification, the adiabatic response coefficients can be expressed by
\begin{equation}
\begin{aligned}
    \mathcal{R}^{\mu\nu} =& - \frac{1}{T} \int_0^{\infty} ds 
    \left\langle 
    \left. \left(
        F_{t}^{\nu} - f_{eq,t}^{\nu}
    \right) \right| \exp{[(i \mathrm{H}_t + \mathrm{D}_t^{\ddag})s]} F_{t}^{\mu} \right\rangle \, . \label{eq:b5}
\end{aligned}
\end{equation}
Here $\delta \rho_t$ is the first order of $1/\tau$,
\begin{equation}
    \delta \rho_t = \int_0^1 dx \rho_{eq, t}^{1-x} \xi_t \rho_{eq, t}^{x} \, , \label{eq:b6}
\end{equation}
and $\langle X | Y \rangle$ represents the scalar product between to superoperator $X$ and $Y$
\begin{equation}
    \langle X | Y \rangle \equiv \int_0^1 dx 
    \text{Tr} \left[
                \rho_{eq,t}^{1-x} X^{\dag}
                \rho_{eq,t}^{x} Y^{\dag}
            \right] \label{eq:b7} \, .
\end{equation}
And the corresponding instantaneous thermodynamic forces are expressed as
\begin{align}
    & F_{t}^{\lambda} \equiv -\partial_{\lambda} H_{\lambda} \\ \nonumber 
    & F_{t}^{T} \equiv -\ln \rho_{eq, t} \label{eq:b8} \, .
\end{align}

Based on the above derivation, we can obtain the geometric properties of the single-qubit heat engine.

Firstly, we notice that the Hamiltonian (\ref{eq:15}) can be represented by the eigenstates $|\phi^{+}\rangle$ and $|\phi^{-}\rangle$ of the energy,
\begin{equation}
    H_{\lambda} = \frac{\hbar\Omega\lambda}{2}
    \left(
    |\phi^{+}\rangle \langle \phi^{+}| - 
    |\phi^{-}\rangle \langle \phi^{-}|
    \right) \, , \label{eq:b9}
\end{equation}
where
\begin{align}
    &|\phi^{+}\rangle = 
    {\left(\begin{array}{ccc}
    \sin{[\theta/2]} \\
    -\cos{[\theta/2]}
    \end{array} \right)} ~~\text{and}~~
    |\phi^{-}\rangle = 
    {\left(\begin{array}{ccc}
    \cos{[\theta/2]} \\
    \sin{[\theta/2]}
    \end{array} \right)} ~~\text{with}~~ \nonumber \\
    &\tan[\theta/2] = \frac{\varepsilon}{\sqrt{\lambda^2-\epsilon^2}+\lambda} \, .  \label{eq:b10}
\end{align}

Thus, according to the quantum detailed balance condition,
($[H_{\lambda}, V^{\pm}] = \pm \hbar\Omega\lambda V^{\pm}$, and ${\rm Tr}\left[V^{\pm}V^{\pm\dag}\right]=\pm\Gamma\Omega\lambda/\left[1-\exp{\left[\mp\hbar\Omega\lambda/T\right]}\right]$), the expression of jump operator $V^{\pm}$ can be obtained,
\begin{equation}
    V^{\pm} = \sqrt{\frac{\pm\Gamma\Omega\lambda}{1-\exp{\left[\mp\hbar\Omega\lambda/T\right]}}} |\phi^{\pm}\rangle \langle \phi^{\mp}| \, . \label{eq:b11}
\end{equation}

As the consequence of quantum detailed balance condition, superoperators $\mathrm{H}_t$ and $\mathrm{D}_t^{\ddag}$ (the adjoint of $\mathrm{D}_t$ with respect to Hilbert-Schmidt scalar product) share the normalized eigenvectors $M^j$ with real eigenvalues $\omega^j$ and non-positive $\kappa^j$ respectively,
\begin{align}
    \mathrm{H}_t M^j = \omega^j M^j \, , ~
    \mathrm{D}_t^{\ddag} M^j = \kappa^j M^j \, , ~
    \langle M^j | M^k \rangle = \delta_{jk} \, . \label{eq:b12}
\end{align}
$\kappa^0 = 0$ and it is the maximum eigenvalue, which corresponds to the unique stationary eigenvector $M^0 = \mathbf{1}$.

These the normalized eigenvectors can be calculated as,
\begin{align}
    M^0 &= \mathbf{1} \, , \nonumber\\
    M^z &= \frac{2}{\hbar\Omega\lambda}\cosh{\left[\hbar\Omega\lambda/2T\right]} H_{\lambda} + \sinh{\left[\hbar\Omega\lambda/2T\right]}\mathbf{1} \,  ~~\text{and}~ \nonumber\\
    M^{\pm} &=\sqrt{\frac{\hbar\Omega\lambda}{T\tanh{\left[\hbar\Omega\lambda/2T\right]}}}|\phi^{\pm}\rangle \langle \phi^{\mp}| \, . \label{eq:b13}
\end{align}
with corresponding eigenvalues,
\begin{align}
    &\kappa^0 = 0, ~~ \kappa^z = -2\Gamma\Omega\lambda\coth{\left[\hbar\Omega\lambda/2T\right]}, ~~ \kappa^{\pm}=\kappa^z/2 \, , \nonumber \\
    &\omega^0 = \omega^z = 0, ~~ \omega^{\pm}=\pm\Omega\lambda \, . \label{eq:b14}
\end{align}
Hence, variances of thermodynamic forces can be calculated. For simplicity, we first calculate the generalized pressure. Notice that 
\begin{equation}
\begin{aligned}
    F_{t}^{\lambda} =&  -\partial_{\lambda} H_{\lambda} \\
    =& -\frac{\hbar\Omega}{2}
    \left(
    |\phi^{+}\rangle \langle \phi^{+}| - 
    |\phi^{-}\rangle \langle \phi^{-}|
    \right) - 
    \frac{\hbar\Omega\lambda}{2} \frac{\partial}{\partial \lambda}
    \left(
    |\phi^{+}\rangle \langle \phi^{+}| - 
    |\phi^{-}\rangle \langle \phi^{-}|
    \right) \\
    =& -\frac{\hbar\Omega}{2}
    \left(
    |\phi^{+}\rangle \langle \phi^{+}| - 
    |\phi^{-}\rangle \langle \phi^{-}|
    \right) - 
    \frac{\hbar\Omega\lambda}{2} \\
    &\left(
    \frac{\partial |\phi^{+}\rangle}{\partial \lambda} \langle \phi^{+}| + |\phi^{+}\rangle \frac{\partial \langle \phi^{+}|}{\partial \lambda} - 
    \frac{\partial |\phi^{-}\rangle}{\partial \lambda} \langle \phi^{-}| - |\phi^{-}\rangle \frac{\partial \langle \phi^{-}|}{\partial \lambda}
    \right) \\
    =& -\frac{\hbar\Omega}{2}
    \left(
    |\phi^{+}\rangle \langle \phi^{+}| - 
    |\phi^{-}\rangle \langle \phi^{-}|
    \right) - 
    \frac{\hbar\Omega\lambda}{2} \frac{\partial \theta}{\partial \lambda}
    \left(
    |\phi^{+}\rangle \langle \phi^{-}| + 
    |\phi^{-}\rangle \langle \phi^{+}|
    \right) \\
    =& -\frac{\hbar\Omega}{2}
    \left(
    |\phi^{+}\rangle \langle \phi^{+}| - 
    |\phi^{-}\rangle \langle \phi^{-}|
    \right) + 
    \frac{\hbar\Omega}{2} \frac{\varepsilon}{\sqrt{\lambda^2-\varepsilon^2}}
    \left(
    |\phi^{+}\rangle \langle \phi^{-}| + 
    |\phi^{-}\rangle \langle \phi^{+}|
    \right) \\
    =& 
    \left(-\frac{\hbar\Omega}{2\cosh\left[\hbar\Omega\lambda/2T\right]} M^z + 
    \frac{\hbar\Omega\tanh\left[\hbar\Omega\lambda/2T\right]}{2} \mathbf{1}
    \right) + \\
    &\frac{\varepsilon\sqrt{\hbar\Omega\lambda
    T\tanh\left[\hbar\Omega\lambda/2T\right]}}
    {2\lambda\sqrt{\lambda^2-\varepsilon^2}}
    \left(M^{+} + M^{-}\right) \, . \label{eq:b15}
\end{aligned}
\end{equation}
And the generalized pressure at instantaneous equilibrium state can be directly calculated from according to the Helmholtz free energy $F =-T \ln Z$, i.e.,
\begin{equation}
    f_{eq,t}^{\lambda} = -\partial_{\lambda} F = \frac{\hbar\Omega\tanh\left[\hbar\Omega\lambda/2T\right]}{2} \, . \label{eq:b16}
\end{equation}
Thus, combining (\ref{eq:b15}) and (\ref{eq:b16}), we obtain the variance of the generalized pressure
\begin{equation}
    F_{t}^{\lambda} - f_{eq,t}^{\lambda} = -\frac{\hbar\Omega}{2\cosh\left[\hbar\Omega\lambda/2T\right]} M^z + 
    \frac{\varepsilon\sqrt{\hbar\Omega\lambda
    T\tanh\left[\hbar\Omega\lambda/2T\right]}}
    {2\lambda\sqrt{\lambda^2-\varepsilon^2}}
    \left(M^{+} + M^{-}\right) \, . \label{eq:b17}
\end{equation}

For the entropy, because
\begin{equation}
\begin{aligned}
    F_t^{T} =& - \ln \rho_{eq,t} \\
    =& \frac{\hbar\Omega\lambda}{2T\cosh\left[\hbar\Omega\lambda/2T\right]} M^z - \frac{\hbar\Omega\lambda\tanh\left[\hbar\Omega\lambda/2T\right]}{2T} + 
    \ln \left[ 2 \cosh\left[\hbar\Omega\lambda/2T\right]\right]  \, , \label{eq:b18}
\end{aligned}
\end{equation}
and the entropy at instantaneous equilibrium state 
\begin{equation}
\begin{aligned}
    f_{eq,t}^{T} =& -\partial_{T} F \\
    =& \ln \left[ 2 \cosh\left[\hbar\Omega\lambda/2T\right]\right] - 
    \frac{\hbar\Omega\lambda\tanh\left[\hbar\Omega\lambda/2T\right]}{2T} \, , \label{eq:b19}
\end{aligned}
\end{equation}
one can derive the variance of entropy as
\begin{equation}
\begin{aligned}
    F_{eq,t}^{T}-f_{eq,t}^{T} =& \frac{\hbar\Omega\lambda}{2T\cosh\left[\hbar\Omega\lambda/2T\right]} M^z \, . \label{eq:b20}
\end{aligned}
\end{equation}

Inserting the above variances of thermodynamic forces into (\ref{eq:b5}), after integrating over $s$, one can obtain the adiabatic response matrix
\begin{eqnarray}
    \mathcal{R} &&= 
    \mathcal{D} + \mathcal{C} = \left(\begin{matrix} 
    \mathcal{D}^{TT} & 
    \mathcal{D}^{T\lambda} \\ 
    \mathcal{D}^{\lambda T} & \mathcal{D}^{\lambda\lambda}
    \end{matrix}\right)  
    + \left(\begin{matrix} 
    0 & 0 
    \\0 & \mathcal{C}^{\lambda \lambda}
    \end{matrix}\right) \nonumber \\
    && = \frac{R}{T^2} 
    \left(\begin{matrix} 
    -\lambda/T & 1 \\ 
    1 & -T/\lambda 
    \end{matrix}\right) + 
    \left(\begin{matrix} 
    0 & 0 \\
    0 & \mathcal{C}^{\lambda \lambda}
    \end{matrix}\right) \, , \label{eq:b21}
\end{eqnarray}
where,
\begin{eqnarray}
    R &&= 
    \frac{\hbar^2\Omega}{8\Gamma}\frac{\tanh{[\hbar\Omega\lambda/2T]}}{\cosh^2{[\hbar\Omega\lambda/2T]}} \\
    \text{and} && \quad \mathcal{C}^{\lambda \lambda} = -\frac{\hbar\varepsilon^2}{2\lambda^2(\lambda^2-\varepsilon^2)}\frac{\Gamma}{\Gamma^2\coth^2{[\hbar\Omega\lambda/2T]}+1} \, . \label{eq:b22}
\end{eqnarray}

\section*{Appendix C: Genetic Algorithm}
In this section, we briefly introduce how genetic algorithm works.

Here, we explain the optimization procedure using a simple maximization problem of one scalar function $f$ in terms of vector $\bm{x}$ in a given domain $D$, i.e.,
\begin{equation}\label{eq:c1}
    \begin{aligned}
        \max_{\bm{x} \in D} f(\bm{x}).
    \end{aligned}
\end{equation}

Firstly, GA requires us to regard every vector $\bm{x}$ in the domain $D$ as a phenotype of an individual. According to the genetic mechanism, a phenotype is uniquely determined by its genotype which can be understood as the gene sequence in the chromosome. GA directly operates the genotype instead of phenotype. The process of mapping the phenotype to a corresponding unique genotype is called parameter coding. Generally, one can code $\bm{x}$ by its binary notation $\bm{x}^b$. The length of $\bm{x}^b$ can be adjusted by a factitiously determined precision $\Delta$ of $\bm{x}$. Because genotype determines the phenotype, once the optimal solution is achieved, the corresponding phenotype of it can be obtained by the decoding process.

Then, the initial population can be presented by a set of chromosomes with genotypes we defined above. For a population of $N$ individuals, we can express it as $G = \{\bm{x}_1^b, \bm{x}_2^b, ..., \bm{x}_N^b\}$. Here these $N$ individuals are randomly selected in the domain $D$.

Next, as illustrated before, not all the individuals in the population can survive, because the environment will obsolete the ones who cannot adapt. The standard is the fitness function. Generally, the objective function $f(\bm{x})$ is chosen to be the fitness function. If one has a low value of fitness, it may die out with a larger probability than others. In this step, $N_d$ individuals will die out.

To ensure the continuation of the population, GA assumes that the surviving individuals can reproduce their off-springs to make up the dying-out ones. It also assumes that any two of them can randomly reproduce. In the reproduction process, generally, the genotype on the chromosome of a child inherits that of either part of their parents randomly, which is called gene crossover. But we also allow genetic mutation to happen. In this situation, some random genes on a chromosome will have a small probability to change randomly. In our example, random sites of the binary notation $\bm{x}^b$ can randomly change for $0$ to $1$ or $1$ to $0$. Crossover and mutation provide the possibility for the child to obtain entirely new genotypes which may have higher fitness than any of their parents.

Thus, the remaining $N-N_d$ parents and the new $N_d$ child construct the new population. Then the selection from the environment happens again. After enough such iterations, the process of natural selection ensures that the fitness of each individual in the population can be optimized. Then, the corresponding optimal genotype can be found.

\section*{Appendix D: Penalty Method}
We describe the basic principles of the penalty method in this section.

The penalty method is applied to deal with the optimization problem with constraints in computer science \cite{yeniay2005penalty,homaifar1994constrained,bertsekas2014constrained}. Note that one can transform any optimization problems with constrains into the minimization problem below,
\begin{equation}\label{eq:d1}
    \begin{aligned}
        \min_{\bm{x}} \quad & f(\bm{x}) &\\
            \text{s.t.} \quad & g_i(\bm{x}) \leq 0 &\quad \quad i = 1,2,...n; \\
            ~ & h_j(\bm{x}) = 0 & ~~ j = 1,2,...m.
    \end{aligned}
\end{equation}
Here $g_i(\bm{x})$ and $h_j(\bm{x})$ is the $i$-th inequality and the $j$-th equality constraint respectively. 

The main idea to deal with this kind of problem is to construct an auxiliary function by the objective and constraint functions so that the original problem can be transformed into one with no constraints. Current methods provide a bunch of theories and algorithms for such auxiliary functions, but the penalty method is one of the most straightforward and efficient tools in numerical calculations. 

The basic starting point of this method is that the auxiliary function $F(\bm{x})$ can be divide into the original objective function $f(\bm{x})$ and an additional penalty function $P(\bm{x})$. The latter is based on the constraint functions. Then, the problem (\ref{eq:d1}) becomes,
\begin{eqnarray}\label{eq:d2}
    \min_{\bm{x}} F(\bm{x})=  f(\bm{x}) + P(\bm{x}),
\end{eqnarray}
where 
\begin{eqnarray}\label{eq:d3}
    P(\bm{x}) = \sigma \left(\sum_{i=1}^{n} \left|\max(0, g_i(\bm{x}))\right| + \sum_{j=1}^{m} \left|h_i(\bm{x}) \right|\right).
\end{eqnarray}
$\sigma$ is a large penal constant.

If $\bm{x}$ is in the feasible region determined by the constraints, $F(\bm{x}) = f(\bm{x})$, i.e., $P(\bm{x}) = 0$. Otherwise, $P(\bm{x})$ will be a sufficiently large number to "punish" the current $\bm{x}$ and let it closer to the feasible region in the next iteration.

\section*{Appendix E: Numerical Setup}
In this section, we provide the details of the hyperparameters and numerical setups. 

In our numerical calculation, we set the range $[0,2]$ of $x$ has been divided into $n_c$ parts. The penalty constant here $\sigma$ is set to be $10^5$. In the parameter-coding step, the precision $\Delta$ is designed to be $10^{-7}$. The size of the initial population $N$ equals $1000$ here. In the selection step, the Roulette wheel method is applied. By normalizing the fitness function of each individual into a positive number, the surviving probability can be calculated from the normalized fitness function divided by the sum of all. Then, $500$ of individuals are selected to be parents and start to reproduce $500$ child. In the gene crossover step, two random sites are selected to be the crossover points on both parents' chromosomes and each fragment divided by the selected sites has $50\%$ probability to be inherited by the child. Next, in the gene mutation step, each site on the chromosome of each offspring is required to be randomly mutated at a small constant probability $p_m$ which is $10^{-3}$ in our calculations. Therefore, $N_s=500$ parents and their $500$ child compose the new population and follow the algorithm iteratively. The iteration time is set to be $10^4$ here. Values of all hyperparameters mentioned in this section are examined to be sufficient for convergence.

\end{document}